\documentclass[12pt]{article}
\usepackage{a4wide}
\usepackage{amssymb}
\usepackage{amsmath}
\usepackage{graphicx}
\usepackage{mathdots}
\usepackage{slashed}
\usepackage{verbatim}
\usepackage{xcolor}

\usepackage{array}
\usepackage{amsthm}

 \numberwithin{equation}{section}

\begin{document}

\begin{flushright}\footnotesize

\texttt{ICCUB-21-007}
\vspace{0.6cm}
\end{flushright}

\mbox{}
\vspace{0truecm}
\linespread{1.1}


\centerline{\LARGE \bf Thermal correlation functions in CFT}
\medskip

\centerline{\LARGE \bf  and factorization}

\medskip


\vspace{.4cm}

 \centerline{\LARGE \bf }

\vspace{1.5truecm}

\centerline{
  
    { \bf D. Rodriguez-Gomez${}^{a,b}$} \footnote{d.rodriguez.gomez@uniovi.es}
   {\bf and}
    { \bf J. G. Russo ${}^{c,d}$} \footnote{jorge.russo@icrea.cat}}

\vspace{1cm}
\centerline{{\it ${}^a$ Department of Physics, Universidad de Oviedo}} \centerline{{\it C/ Federico Garc\'ia Lorca  18, 33007  Oviedo, Spain}}
\medskip
\centerline{{\it ${}^b$  Instituto Universitario de Ciencias y Tecnolog\'ias Espaciales de Asturias (ICTEA)}}\centerline{{\it C/~de la Independencia 13, 33004 Oviedo, Spain.}}
\medskip
\centerline{{\it ${}^c$ Instituci\'o Catalana de Recerca i Estudis Avan\c{c}ats (ICREA)}} \centerline{{\it Pg.~Lluis Companys, 23, 08010 Barcelona, Spain}}
\medskip
\centerline{{\it ${}^d$ Departament de F\' \i sica Cu\' antica i Astrof\'\i sica and Institut de Ci\`encies del Cosmos}} \centerline{{\it Universitat de Barcelona, Mart\'i Franqu\`es, 1, 08028
Barcelona, Spain }}
\vspace{1cm}

\centerline{\bf ABSTRACT}
\medskip

We study $2$-point and $3$-point functions in CFT at finite temperature for large dimension operators using holography. The 2-point function leads to a universal formula for the holographic free energy in $d$ dimensions in terms of the $c$-anomaly coefficient. 
By including $\alpha'$ corrections to the black brane background, one can reproduce the leading correction at strong coupling. In turn,  3-point functions have a very intricate structure, exhibiting a number of interesting properties. In simple cases, we find an analytic formula, which reduces to the expected expressions in different limits. When the dimensions satisfy $\Delta_i= \Delta_j+ \Delta_k$, the thermal 3-point function satisfies a factorization property. We argue that in $d>2$ factorization is a reflection of the semiclassical regime.

\noindent

\newpage

\tableofcontents

\section{Introduction}

Conformal Field Theories play a pivotal role in Quantum Field Theory as endpoints of  trajectories of the Renormalization Group. 
They also govern the near-critical behavior  of phase transitions occurring in  relevant models of statistical mechanics and in physical theories. 
Introducing finite temperature is equivalent to considering a $d$-dimensional CFT placed in 
euclidean $S^1\times \mathbb{R}^{d-1}$, which implies a periodic identification of the euclidean time, $\tau =\tau+\beta$. The response to this non-trivial geometry provides an extra probe of the structure of the theory beyond the usual quantization on $\mathbb{R}^d$ (or its conformal cousins such as $\mathbb{R}\times S^{d-1}$),
which reveals important aspects of the dynamics of the theory.

If one considers two operators inserted within a ball of radius much smaller than $\beta $ --~so that effectively they see flat space around~-- it is possible to use the zero-temperature OPE.
This was studied in general in \cite{Iliesiu:2018fao} (see also \cite{ElShowk:2011ag}) for the case of the 2-point function, finding an elegant expansion in terms of Gegenbauer polynomials, whose coefficients incorporate both the OPE data as well as the non-zero vacuum expectation values in the thermal background. For strongly coupled theories with a gravity dual this structure naturally arises from the computation of the 2-point function in the black brane background, as recently shown for operators of large dimension
in \cite{Rodriguez-Gomez:2021pfh}.\footnote{See also \cite{Alday:2020eua} for a similar computation in thermal $AdS$.
This corresponds to a boundary theory with no energy-momentum tensor.} It is interesting to note that, through the Eigenstate Thermalization Hypothesis, the thermal 2-point function is related to Heavy-Light-Light-Heavy 4-point correlators at $T=0$ (\textit{e.g.} \cite{Lashkari:2016vgj,Fitzpatrick:2019zqz,Fitzpatrick:2019efk,Karlsson:2021duj}). 

The main object of interest in this paper will be thermal (connected) 3-point functions in holographic theories.
While at $T=0$ the structure of three-point functions is completely fixed by conformal invariance (just like for 2-point functions), 
at finite temperature they have an extremely complicated structure,  involving  multiple regimes.

Holographic $n$-point functions can be computed by means of an expansion in Witten diagrams in the appropriate gravitational
background.
 3-point functions are distinguished from   higher-point functions because there is only one possible Witten diagram, involving a cubic coupling. As a result, the value of the bulk cubic coupling enters only as an overall constant in the final result. 
 Just as in \cite{Rodriguez-Gomez:2021pfh}, we will focus on correlators of operators with large dimension. 
 In this case, the propagator can be computed in the WKB approximation,
 where it is represented as the exponential of the geodesic length
 of the trajectory for a particle moving between the boundary point where the operator is inserted up to the point of interaction.
 At $T=0$, the 3-point function has been computed using this technique in \textit{e.g.} \cite{Klose:2011rm,Buchbinder:2011jr,Minahan:2012fh}. A number of new interesting issues arise when extending the method
 to finite temperature, as we shall discuss.

Thermal correlation functions contain very interesting information about quantum systems including their chaotic nature (see \textit{e.g.} \cite{Dymarsky:2021bjq} for a recent application). In holographic theories, it has been shown that they can probe the interior of the black hole in the gravitational dual (\textit{e.g.} \cite{Kraus:2002iv,Fidkowski:2003nf,Hamilton:2006fh,Balasubramanian:2011ur,Heemskerk:2012mn,Papadodimas:2012aq,Hartman:2013qma,Dodelson:2020lal,Grinberg:2020fdj,Rodriguez-Gomez:2021pfh}). 
In particular, they are expected to shed new light on the nature of spacetime singularities, although this may require going far beyond the leading strong coupling  order.
We leave these studies for future work.

This paper is organized as follows. In section \ref{generalities} we discuss the general set-up for the computation of 3-point functions for operators of large dimension using the geodesic approximation. 
We define {\it ingoing} and {\it returning} geodesics and describe
the different contributions to the saddle-point equations.
Details of the derivations are relegated to appendix \ref{appendix:geodesics}.  In section \ref{2-point} we adapt the discussion to the case of (connected) 2-point functions in $d$-dimensions. By matching with the expected form for the 2-point function in QFT we obtain a formula for the free energy of holographic theories as a function of their central charge $c_{\mathcal{T}}$.
Using the known value of the free energy for holographic theories in \cite{Gubser:1998nz}, we test this formula by matching the central charges in $d=3,4,6$ against the known values read off from \textit{e.g.} \cite{Osborn:1993cr,Bastianelli:1999ab,Chester:2014fya}. For the $d=4$ case, we discuss the leading $\alpha'$ correction to the free energy, reproducing the  results of \cite{Gubser:1998nz}. In section \ref{3-point-symmetric} we initiate the study of 3-point functions, concentrating on a symmetric configuration, for which analytic results can be found. 
In section \ref{3-point} we study the general three-point correlation function of scalar operators. We show that 
in an extremal limit
they satisfy a remarkable factorization property.

\section{Holographic 3-point functions: setting up the stage}\label{generalities}

We are interested on thermal correlation functions of scalar operators in holographic $d$-dimensional CFT's. In particular, we will examine
two-point functions and the three-point function,

\begin{equation}
    \langle O_{\Delta_1}(x_1)\,O_{\Delta_2}(x_2)\,O_{\Delta_3}(x_3)\rangle\,.
\end{equation}
To simplify the the problem, we will mostly focus on the case where operators are inserted at the same spatial point, which, with no loss of generality can be taken to be $\vec{0}$, that is, we will consider $x_i=(t_i,\vec{0})$.

On general grounds, the finite temperature state is dual to a black brane in $AdS_{d+1}$

\begin{equation}
    ds^2=\frac{R^2}{z^2}\Big[-f(z)\,dt^2+\frac{dz^2}{f(z)}+d\vec{x}^2\Big]\,,\qquad f(z)=1-\frac{z^d}{z_0^d}\,  ,\ \ z_0=\frac{d}{4\pi }\, \beta\ .
\end{equation}
We reserve $t$ for the time coordinate with Lorentzian signature and $\tau$ for the Wick-rotated euclidean coordinate.
Each operator is dual to a fluctuating scalar field $\phi_i$ in the black brane in $AdS_{d+1}$ with its mass related by the standard holographic formula to the dimension of the dual operator,
$$
\Delta=\frac{d}{2} +\sqrt{\frac{d^2}{4}+m^2R^2}\ .
$$
We shall consider operators of large scaling dimensions. For such operators,
$\Delta\approx mR$ when $mR\gg 1$.
 
 The bulk lagrangian is then obtained by expanding the corresponding gravitational action in the fluctuations, where terms higher-than-quadratic in the fluctuations --suppressed by $\frac{1}{N}$ in terms of CFT variables-- give rise to bulk Witten diagrams contributing to higher-point correlators in the boundary theory. Note that 3-point functions are somewhat special in that only cubic vertices contribute to them. Hence the relevant part of the  bulk lagrangian looks like

\begin{equation}
\label{bulkS}
    S_{\rm bulk}=\int d^d x \sqrt{g}\left( \frac{1}{2} \partial\phi_i\partial\phi_i+\frac{1}{2}m_i^2\phi_i^2+\lambda_{ijk}\phi_i\phi_j\phi_j\right)\,.
\end{equation}
The $\lambda_{ijk}$ couplings depend on the particular model. However, since such couplings will only appear as  overall constants in front of the 3-point function, their precise value will not be relevant in this paper.

In order to compute the 3-point function holographically starting with \eqref{bulkS} we should evaluate the corresponding 3-leg Witten diagram. Essentially

\begin{equation}
    \langle O_{\Delta_1}(x_1)\,O_{\Delta_2}(x_2)\,O_{\Delta_3}(x_3)\rangle=\lambda_{123}\int du\,G_{\Delta_1}(x_1,u)\,G_{\Delta_2}(x_2,u)\,G_{\Delta_3}(x_3,u)\,,
    \label{uuii}
\end{equation}
where $G_{\Delta_i}(x_i,u)$ is the boundary-to-bulk propagator for a field dual to an operator of dimension $\Delta$ from $x_i$ in the boundary to the bulk point $u=(\vec{x},z)$. This propagator comes from solving, with the appropriate boundary conditions, the Klein-Gordon equation in the black brane background. Such computation dramatically simplifies in the case of $\Delta_i\gg 1$. In that case one can approximate $G_{\Delta}(x,u)$ by the WKB approximation, which leads to

\begin{equation}
    G_{\Delta}(x,u)=e^{-iS_{\Delta}(x,u)}=e^{-i\Delta\,\ell(x,u)}\,,
\end{equation}
being $S_{\Delta}(x,u)$ the action for a particle of mass $m\sim R^{-1}\Delta$ which departs from the boundary at $x$ and arrives to the bulk point $u$. Alternatively, this can be thought of as the length $\ell(x,u)$ of a geodesic from $x$ to $u$. Thus, for $\Delta_i\gg 1$ (but still, $\Delta_i\ll c_{\mathcal{T}}$), the three-point correlation function of interest can be computed as

\begin{equation}
    \langle O_{\Delta_1}(x_1)\,O_{\Delta_2}(x_2)\,O_{\Delta_3}(x_3)\rangle=\lambda_{123}\int du\,e^{-i\Delta_1\,\ell(x_1,u)}\,e^{-i\Delta_2\,\ell(x_2,u)}\,e^{-i\Delta_2\,\ell(x_2,u)}\, ,
\end{equation}
where $u$ now represents the intersecting point of the three geodesic arcs.
Under the assumption $\Delta_i\gg 1$, the integral over the bulk point $u$ is determined by a  saddle-point calculation. Thus, one finds

\begin{equation}
\label{finalformula}
    \langle O_{\Delta_1}(x_1)\,O_{\Delta_2}(x_2)\,O_{\Delta_3}(x_3)\rangle\sim e^{-iS_{\rm os}}\,,
\end{equation}
where $S_{\rm os}$ stands for the ``action" $S=\Delta_1\ell(x_1,u_I)+\Delta_2\ell(x_2,u_I)+\Delta_3\ell(x_3,u_I)$ evaluated on-shell --i.e. at the saddle-point $u_I$,   solving the equation $dS/du=0$.

\bigskip

Our first task is to compute the boundary-to-bulk propagator in the black brane background, a.k.a. the geodesic arcs for the case of $\Delta\gg 1$. Since we consider all operators inserted at $\vec{0}$, it is clear that the spatial part of $u_I$ at the saddle point will be at the origin. We can thus focus on geodesics which move in the plane $(z,t)$. 
The action for a massive particle moving along $t$ whose worldline is parametrized by $z$ is given by

\begin{equation}
    S=- \Delta\,\int dz\,\frac{1}{z}\,\sqrt{f\,\dot{t}^2-\frac{1}{f}}\, ,
\end{equation}
where $\dot t=dt/dz$. Since the action does not depend on $t$, the corresponding conjugate momentum --~which  will be denoted as $p_t=\Delta\,\mu$~-- is conserved. This gives a first-order equation

\begin{equation}
\label{dott}
  \dot \tau = i  \dot{t}= \pm \frac{ z\,\mu}{f\,\sqrt{f- \mu^2\,z^2}}\,,
\end{equation}
where we have introduced the euclidean time $\tau=i t$. The signs $\pm $ in \eqref{dott} respectively  describe either a geodesic going towards the black hole or a geodesic going towards the boundary.

The parameter $\mu $ can be viewed as an integration constant, which is determined by demanding that the geodesic departs from $(\tau=\tau_1,z=0)$ and reaches the intersection point $(\tau_I, z_I)$. At this point,
\begin{equation}
    \tau_I =\tau_1+ \int_0^{z_I} dz\,\frac{z\,\mu}{f(z)\,\sqrt{f(z)-\mu^2\,z^2}}\,.
\label{tamp}
\end{equation}
This can be viewed as a condition that implicitly defines $\mu=\mu(\tau_I,z_I)$. The equation \eqref{tamp} cannot be inverted explicitly to find an explicit formula for $\mu(\tau_I,z_I)$, except in the case of $d=2$.
Nevertheless, the implicit definition implied by the condition \eqref{tamp} is enough in order to determine the saddle point and thereby the three-point correlation function.

\begin{figure}[h!]
    \centering
    \includegraphics[scale=.5]{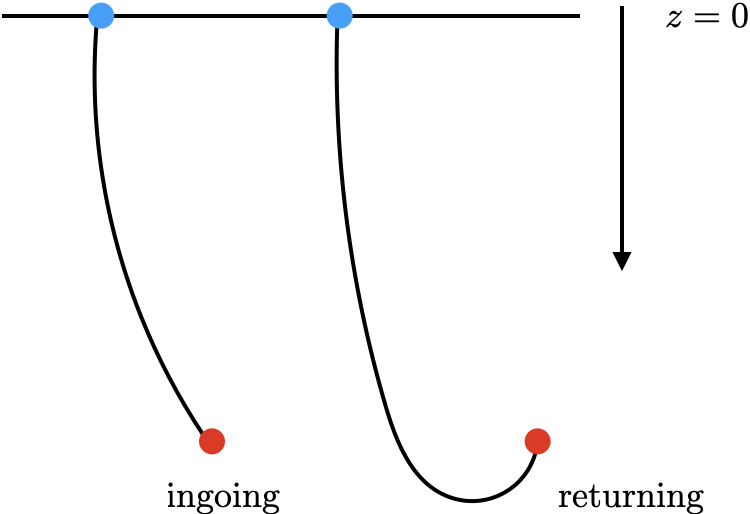}
    \caption{The two types of geodesics to consider.}
    \label{geodesicarcs}
\end{figure}

From \eqref{dott}, it is clear that the maximum reach inside the bulk of the geodesic,
{\it i.e.} the turning point where $dz/d\tau=0$, is given by the equation

\begin{equation}
\label{zmax}
    f(z_{\rm max})=\mu^2\,z_{\rm max}^2\,.
\end{equation}
In \eqref{tamp} we assumed that the sign describing the geodesic going towards the black hole. These will be called \textit{ingoing} geodesics, that is, geodesics departing from the boundary at $x$ and arriving to $u_I$ before getting to $z_{\rm max}$. For the calculation of the three-point function, configurations where geodesics reach $z_{\rm max}$ and then
 meet $u_I$ in their return way will also be relevant. These will be called \textit{returning} geodesics.
These two types of geodesics are shown in fig.\eqref{geodesicarcs}.

The analysis of each case is carried out in appendix \ref{appendix:geodesics}. 
The contribution to the action for each type of geodesic is, respectively, given by

\begin{equation}
\label{Singoing}
    S_{\rm ingoing}= -i \Delta\,\int_{0}^{z_i}dz\,\frac{1}{z\,\sqrt{f- \mu^2\,z^2}}\,.
\end{equation}
and 

\begin{equation}
\label{Sreturning}
    S_{\rm returning}=-i\Delta\,\int_{0}^{z_{\rm max}}dz\,\frac{1}{z\,\sqrt{f(z)-\mu^2\,z^2}}+i\Delta\,\int_{z_{\rm max}}^{z_I}dz\,\frac{1}{z\,\sqrt{f(z)-\mu^2\,z^2}}\,.
\end{equation}
As discussed above, here $\mu$ is to be understood as an integration constant, which, through the equation of motion (see appendix \ref{appendix:geodesics}), has to be tuned so that the geodesic arc meets the desired boundary conditions: departing the boundary from the corresponding point $x$ and arriving at the intersection point $u_I$.

The total action in \eqref{finalformula} will be a sum of geodesic arcs, each of them either ingoing --contributing as \eqref{Singoing}-- or returning --contributing as \eqref{Sreturning}. The choice of ingoing/returning is determined by the dynamics.

The formula \eqref{finalformula} for the three-point function  requires to evaluate the $u$ integral through the saddle-point method, which implies an extremization of the total action in the $\tau $ and $z$ direction.
Using the formulas of appendix \ref{appendix:geodesics}, we obtain

\begin{eqnarray}
\label{saddleequations}
    &&\frac{dS}{dz_I}=0\ :\qquad \sum_{i=1}^3 \epsilon_i \Delta_i\,\sqrt{f(z_I)-\mu_i^2\,z_I^2}=0\ ,
    \nonumber\\
     &&\frac{dS}{d\tau_I}=0\ :\qquad \sum_{i=1}^3 \Delta_i \mu_i =0\ .
\end{eqnarray}
where $\epsilon =1$ for  ingoing geodesics and $\epsilon =-1$ for  returning geodesics.

\section{Thermal 2-point functions in $d$ dimensions}\label{2-point}

The 2-point functions of operators with $\Delta\gg 1$ can be computed in terms
of  a single geodesic departing from the boundary from a point $x_1$, making an excursion into the bulk  and coming back to another point $-x_1$. One writes
\begin{equation}
\label{finalformula2-point}
    \langle O_{\Delta}(x_1)\,O_{\Delta_2}(-x_1)\rangle\approx e^{-iS}\,.
\end{equation}
The geodesic is uniquely determined in terms of $x_1$.
Alternatively, the two-point function can also be computed using the formalism developed above.
We view the geodesic as two geodesic arcs joining at some bulk point $u$  (see \textit{e.g.} \cite{Dobashi:2002ar,Dobashi:2004nm,Janik:2010gc}).
Then the two-point function can be computed as the integral 
\begin{equation}
   \langle O_{\Delta}(x_1)\,O_{\Delta_2}(-x_1)\rangle\approx 
   \int du\,e^{-i\Delta\,\ell(x_1,u)}\,e^{-i\Delta\,\ell(-x_1,u)}\, .
\end{equation}
The integral can be evaluated by a saddle-point approximation, with the action being 
 $S=\Delta\ell(x_1,u)+\Delta\ell(x_2,u)$. By symmetry,
 the saddle-point must be located at $(\vec 0,z_I)$. Using \eqref{saddleequations}, the saddle point equations are

\begin{equation}
\Delta\,\mu_1+\Delta\,\mu_2=0\,,\qquad \epsilon_1\Delta \sqrt{f(z_I)-\mu_1^2\,z_I^2}+\epsilon_2\Delta\sqrt{f(z_I)-\mu_2^2\,z_I^2}=0\,.
\end{equation}
The first condition sets $\mu_1=-\mu_2=\mu$ (which is nothing but momentum conservation). Since both geodesics are ingoing,  $\epsilon_1=\epsilon_2=1$.
Thus we get the condition

\begin{equation}
\label{condmax}
    f(z_I)-\mu^2\,z_I^2=0\,.
\end{equation}
This is precisely the maximum reach for the geodesic, see  \eqref{zmax},
which implies that the arcs are just the arms of a single U-shaped geodesic, as expected, in this way reproducing the result obtained by a direct calculation
of the action \eqref{finalformula2-point}.

 Let us now consider the explicit calculation of the correlator $\langle O_{\Delta}(\tau_1)O_{\Delta}(-\tau_1)\rangle$ in $d$ dimensions. It is convenient to solve \eqref{condmax} for $\mu $, so that $\mu^2=\frac{f(z_I)}{z_I^2}$. Then, using the formulas from appendix \ref{appendix:geodesics}, we have 
 
 \begin{eqnarray}
    &&S=-2i\,\Delta z_I\,\int_{0}^{z_I}dz\,\frac{1}{z\,\sqrt{f(z)\,z_I^2-f(z_I)\,u^2}}\,,\nonumber \\ 
    &&\tau_I=\tau_1+\sqrt{f(z_I)}\int_0^{z_I} dz\,\frac{z}{f(z)\,\sqrt{f(z)\,z_I^2-f(z_I)\,z^2}}\,.
    \label{trenta}
\end{eqnarray}
The correlation function is found by substituting into ${\rm exp}(-i S)$ the expression of $z_I$ in terms of $\tau_I$, implicitly defined  by \eqref{trenta}. 
Since \eqref{trenta} cannot be inverted in general, we can work in perturbation theory in the temperature. This is equivalent to a short-distance expansion of the correlator and it amounts to expanding \eqref{trenta} in powers of $z^2$.
We find 

\begin{equation}
    z_I=\tau_1+\frac{\tau_1^{d+1}}{4\,z_0^d}\,\Big(\frac{(d-1)\,\sqrt{\pi}\,\Gamma(1+\frac{d}{2})}{\Gamma(\frac{d+3}{2})}-2\Big)+\cdots\,.
\end{equation}
This gives

\begin{equation}
\label{thermal2point}
    \langle O(\tau)\,O(0)\rangle=\frac{1}{\tau^{2\Delta}}\ \exp \left[ 
  \alpha_d\,\Delta\,T^d\,\tau^d\,(1+\mathcal{O}\big((T\tau)^d)\big) \right]\, ,\qquad \alpha_d \equiv  \frac{\pi^{d+1}\,\Gamma(d)}{d^d\,\Gamma(\frac{d+3}{2})\,\Gamma(\frac{d-1}{2})}\,.
\end{equation}
where  $\tau=\frac{\tau_1}{2}$.

\subsection{Universal formula for the free energy}

The thermal 2-point function, in the regime $T|x|\ll 1$, contains interesting information about the $O\times O$ OPE  

\begin{equation}
\label{OPE}
 O(x)O(0)= \sum_{\mathcal{O}}\frac{f_{OO\mathcal{O}}}{c_{\mathcal{O}}}\, \frac{x_{\nu_1}\cdots x_{\nu_{J_{\mathcal{O}}}}}{|x|^{2\Delta_O-\Delta_{\mathcal{O}}+J_{\mathcal{O}}}}\,\mathcal{O}^{\nu_1\cdots\nu_{J_{\mathcal{O}}}}(0)+{\rm descendants}\,.
 \end{equation}
where $c_{\mathcal{O}}$ is the coefficient of the 2-point function of the primaries $\mathcal{O}$ normalized as

\begin{equation}
\label{2point}
\langle \mathcal{O}^{\mu_1\cdots\mu_{J_{\mathcal{O}}}}(x)\,\mathcal{O}^{\nu_1\cdots\nu_{J_{\mathcal{O}}}}(0)\rangle=c_{\mathcal{O}}\,\frac{I^{\mu_1\nu_1}\cdots I^{\mu_{J_{\mathcal{O}}}\nu_{J_{\mathcal{O}}}}-{\rm traces}}{|x|^{2\Delta_{\mathcal{O}}}}\,,\qquad I^{\mu\nu}=\delta^{\mu\nu}-2\frac{x^{\mu}x^{\nu}}{|x|^2}\,.
\end{equation}
The key point is that the $S^1\times \mathbb{R}^{d-1}$ geometry provides, through the circle, both a special direction and a scale allowing for non-zero VEVs of the form,

\begin{equation}
    \langle \mathcal{O}^{\mu_1\cdots\mu_{J_{\mathcal{O}}}}(x)\rangle = \frac{b_\mathcal{O}}{\beta^{\Delta_{\mathcal{O}}}}\,(e^{\mu_1\cdots\mu_{J_{\mathcal{O}}}}-{\rm traces})\,,\qquad e^{\mu}=\delta^{\mu}_{\tau}\,.
\end{equation}
Combining with \eqref{OPE} and doing the tensor contractions, one finds that, when $T|x|\ll 1$, the thermal 2-point function  admits the expansion \cite{Iliesiu:2018fao}

\begin{equation}
\label{thermm}
 \langle O(x)O(0)\rangle=\sum_{\mathcal{O}}  \frac{a_{OO\mathcal{O}}}{\beta^{\Delta_{\mathcal{O}}}}\, \frac{1}{|x|^{2\Delta_O-\Delta_{\mathcal{O}}}}\,C_{J_{\mathcal{O}}}^{(\nu)}(\eta)\,,\qquad a_{OO\mathcal{O}}=\frac{f_{OO\mathcal{O}}\,b_{\mathcal{O}}}{c_{\mathcal{O}} }\, \frac{J_{\mathcal{O}}!}{2^{J_{\mathcal{O}}}\,(\nu)_{J_{\mathcal{O}}}}\, ,
 \end{equation}
where $\eta\equiv \tau/|x|$ and $C_{J_{\mathcal{O}}}^{(\nu)}(\eta)$ are Gegenbauer polynomials ($\nu\equiv \frac{d-2}{2}$).

As noted in \cite{Rodriguez-Gomez:2021pfh} for the $d=2,4$ cases, the thermal 2-point function \eqref{thermal2point}
becomes exact in the double-scaling limit,

\begin{equation}
    \label{thermallimit}
    \Delta\rightarrow \infty,\,\qquad |x|T\rightarrow 0\,,\qquad \Delta\,(|x|T)^d={\rm fixed}\, .
\end{equation}
In this limit, the $\mathcal{O}\big((T\tau)^d\big)$ terms in \eqref{thermal2point} vanish. 
The full spacetime dependence can now be restored using the consistency with the OPE
\eqref{thermm}. This implies that the exponent must be the $\vec{x}=0$ limit of the second Gegenbauer polynomial (this was shown explicitly to be the case in $d=2,4$ in \cite{Rodriguez-Gomez:2021pfh}). Therefore we can write

\begin{equation}
\label{thermalppp}
    \langle O(\tau)\,O(0)\rangle=\frac{1}{\tau^{2\Delta}}\ \exp \left[
  A_d\,\Delta\,T^d\,|x|^d\,C_2^{(\frac{d-2}{2})}(\eta) \right]\,, \qquad A_d \equiv  \frac{2^{d-2}\,\pi^{\frac{1+2d}{2}}\,\Gamma\Big(\frac{d-2}{2}\Big)}{d^d\,\Gamma\Big(\frac{d+3}{2}\Big)}\,.
  \end{equation}

Let us consider explicitly the case of $d=4$. Expanding the exponential, one finds

\begin{equation}
\langle O_{\Delta}(x)O_{\Delta}(0)\rangle=\sum_{J=0}\,\sum_{n=0} \,\frac{1}{n!}\,\Big(\frac{\pi^4\,\Delta}{120}\Big)^n\,\frac{\mathcal{C}_{J,n}}{\beta^{4n}}\,\frac{1}{|x|^{2\Delta-4n}}\,C_J^{(1)}(\eta)\,,
\end{equation}
where the numerical coefficients $\mathcal{C}_{J,n}$ turn out to be given as well in terms of Gegenbauer polynomials as

\begin{equation}
\label{Cs}
\mathcal{C}_{J,n}=\frac{2}{\pi}\,\int_{-1}^1d\eta\,\sqrt{1-\eta^2}\,C_J^{(1)}(\eta)\,\big(C_2^{(1)}(\eta)\big)^n=\begin{cases}C_{n-\frac{J}{2}}^{(-n)}(-\frac{1}{2})-C_{n-\frac{J}{2}-1}^{(-n)}(-\frac{1}{2}), \quad 
&{\rm if}\ n,\,J\ne 0\\ 1\quad &{\rm if}\ n,\,J=0\end{cases}
\end{equation}
Comparing with the general form of the correlator, each term in the series in \eqref{thermalppp} represents the contribution of a spin $J$ and dimension $4n$ operator constructed as the appropriate contraction of $n$ factors of energy-momentum tensors, which we will denote schematically as $\mathcal{T}^n$. Clearly, the spin of such contraction is at most $2n$. Consistently, the coefficients in \eqref{Cs} vanish if $J>2n$. Moreover, the contribution of each term scales with $\Delta^n=\Delta^{\frac{\Delta_{\mathcal{T}^n}}{4}}$, which is precisely the large $\Delta$ scaling expected \cite{Lashkari:2016vgj} (see also \cite{Gobeil:2018fzy}). Note that this extends to arbitrary dimensions, as the expansion of \eqref{thermalppp} involves $\Delta^n=\Delta^{\frac{dn}{d}}=\Delta^{\frac{\Delta_{\mathcal{T}^n}}{d}}$.

The leading term in the expansion of  \eqref{thermalppp} corresponds to the $J=2$ term in \eqref{thermm}. This yields the identification
 $a_{\mathcal{T}}=A_d$. 
 The OPE coefficient $f_{OO\mathcal{T}}$  with the energy-momentum tensor is fixed by a Ward identity. One has \cite{Iliesiu:2018fao}
 \begin{equation}
     f_{OO\mathcal{T}} = -\frac{d}{d-1}\frac{\Delta}{{\rm vol}(S^{d-1})}\ .
 \end{equation}
 Thus, combining with \eqref{thermm}, we find a universal formula for the vacuum expectation value  for the energy density

\begin{equation}
    \frac{\langle \mathcal{T}^{00}\rangle_{\beta}}{c_{\mathcal{T}}}=-\frac{2^{2d-1}\,(d-1)^2\,\pi^{\frac{3d}{2}}\,\Gamma(\frac{d}{2})}{d^d\,\Gamma(d+2)}\,T^d\,.
\end{equation}
This formula can be alternatively derived from the thermodynamics of black holes supplemented with the  AdS/CFT dictionary relating the Newton constant and $c_{\mathcal{T}}$ as in \textit{e.g.} \cite{Kovtun:2008kw}.

Using the standard thermodynamic relation, $F=E+T\,\frac{dF}{dT}$, we find

\begin{equation}
\label{F}
    F=-\frac{2^{2d-1}\,(d-1)\,\pi^{\frac{3d}{2}}\,\Gamma(\frac{d}{2})}{d^d\,\Gamma(d+2)}\,c_{\mathcal{T}}\,T^d\,.
\end{equation}

We can now combine this formula with the general expression for the free energy given in \cite{Gubser:1998nz} (see also \cite{Klebanov:1996un}) to provide an independent derivation of the central charge for ABJM, $\mathcal{N}=4$ SYM and the 6d $(2,0)$ theory for $d=3,4,6$ respectively.
We find

\begin{equation}
    c_{\mathcal{T}}^{\rm ABJM}=\frac{2\sqrt{2}}{\pi^3}\,N^{\frac{3}{2}}\,,\qquad c_{\mathcal{T}}^{\mathcal{N}=4}=\frac{10}{\pi^2}\,N^2\,,\qquad c_{\mathcal{T}}^{(2,0)}=4N^3\ \frac{84}{\pi^6}\,.
\end{equation}
Note that these central charges are in the convention where a free real scalar contributes as

\begin{equation}
    c_{\mathcal{T}}^{\rm free\,scalar}=\frac{d}{d-1}\,\frac{1}{S_d^2}\,,\qquad S_d=\frac{2\pi^{\frac{d}{2}}}{\Gamma\Big(\frac{d}{2}\Big)}\,.
\end{equation}
For $\mathcal{N}=4$ SYM, using \cite{Osborn:1993cr} with $n_{\phi}=6$ real scalars, $n_{\psi}=\frac{4}{2}$ Dirac fermions and one gauge field in $d=4$, one precisely recovers the expression of $c_{\mathcal{T}}^{\mathcal{N}=4}$ given above. 

Consider now ABJM theory. It is convenient to  divide by $c_{\mathcal{T}}^{\rm free\,scalar}$,  so that we find the central charge in the conventions where a free real scalars contributes 1 --we will denote it with a hat. We find 

\begin{equation}
    \hat{c}_{\mathcal{T}}^{ABJM}=\frac{64\sqrt{2}}{3\pi}\,N^{\frac{3}{2}}\,,
\end{equation}
in agreement with \cite{Chester:2014fya} for $k=1$ (see eq. (5.11)) (see also  \cite{Bastianelli:1999ab}). 

Finally, for the 6d $(2,0)$ theory, note that in our conventions, a free tensor multiplet contributes as \cite{Bastianelli:1999ab} (see eqs. (2.6) and (3.23))

\begin{equation}
    c_{\mathcal{T}}^{\rm free\,6d\,tensor}=\frac{84}{\pi^6}\,.
\end{equation}
Thus, in units where the free tensor multiplet contributes 1, we have

\begin{equation}
    \hat{c}_{\mathcal{T}}^{(2,0)}=4N^3\,,
\end{equation}
 reproducing the known result.
 
 
\subsection{Including $\alpha'$ corrections in the $d=4$ case}

 One of the famous problems in AdS/CFT concerns
the match between the holographic free energy and the  free energy derived from ${\cal N}=4$ super Yang-Mills theory. As known, the weak-coupling free energy differs from the strong-coupling (holographic) result by a factor of 3/4.
It is believed that both limits arises from a monotonic  function $f(\lambda)$, which interpolates between 1 and 3/4. The leading strong-coupling correction to $f(\lambda)$ was computed in \cite{Gubser:1998nz} by computing the explicit modifications to the black brane geometry induced by the ${\alpha'}^3 R^4 $ corrections
to the type IIB effective action.
In the present formalism,
we may similarly include corrections  by considering the geodesic moving in the $\alpha'$ corrected background. 
We will ignore  other possible sources of $\alpha'$ corrections, such as possible higher-derivative interactions of the bulk scalar field, assuming that the leading-order correction comes from the $\alpha'$ corrected background.

The first observation is that $\alpha'$  corrections arise in powers of $\frac{z^4}{z_0^4}$. 
Thus the net effect of the $\alpha'$ corrections will be to change the $a_{OO\mathcal{O}}$ coefficients, as opposed to turning on new terms (for instance, a term proportional to $C_0^{(1)}$ corresponding to a dimension 2 scalar is not induced, as it would require a $\frac{z^2}{z_0^2}$ correction). As a result of corrections to the background, the relation between $z_0$ and the temperature also gets $\alpha'$ corrections. It becomes

\begin{equation}
\label{alphaprimecorrectedz0}
    z_0=\frac{1+15\gamma}{\pi T}\, , \qquad \gamma = \frac{1}{8}\zeta(3) (\alpha')^3\ ,
\end{equation}
in units where the AdS radius is equal to 1.  
 The correction to the energy density $\langle \mathcal{T}^{00}\rangle_{\beta} $ shows up in the coefficient of the term $T^4$
in the expansion of the exponent in \eqref{thermalppp}. This is affected by the leading $\frac{z^4}{z_0^4}$ term in the expansion.
Therefore, the leading $\alpha'$ correction to the energy density can be obtained from the $O(z^4)$ correction in the metric.  Formally, one gets the same metric as in the unperturbed case, by trading $z_0^4$ by $ z_{0,{\rm corrected}}^4\,(1+75\gamma)^{-1}$, where we have included the subscript to emphasize that this is the $\alpha'$ corrected $z_0$ given in \eqref{alphaprimecorrectedz0}. Using \eqref{alphaprimecorrectedz0}, we obtain the re-scaling,

\begin{equation}
    z_0=\frac{1}{\pi T}\rightarrow \frac{(1+15\gamma)\,(1+75\gamma)^{-\frac{1}{4}}}{\pi T}\sim \frac{1}{\pi T\,(1+15\gamma)^{\frac{1}{4}}}\, .
\end{equation}
Thus, 

\begin{equation}
    T^4\rightarrow T^4\,(1+15\gamma) = T^4\,\left( 1+\frac{15}{8}\zeta(3)\,(2g_{YM}^2N)^{-\frac{3}{2}}\right)\ .
    \end{equation}
This leads to the free energy    
    \begin{equation}
     F   = - \frac{\pi^2}{6}\, N^2 T^4\,\left( \frac{3}{4}+\frac{45}{32}\zeta(3)\,(2g_{YM}^2N)^{-\frac{3}{2}}+\cdots\right)\,
    \,.
\end{equation}
This is precisely the leading correction at strong coupling 
found in \cite{Gubser:1998nz}. 


\section{Thermal 3-point correlation functions\\ with reflection symmetry}\label{3-point-symmetric}

Let us now consider  thermal 3-point functions at coincident spatial points. According to the previous discussion, this is computed holographically as

\begin{equation}
    \langle O_{\Delta_1}(\tau_1)\,O_{\Delta_2}(\tau_1)\,O_{\Delta_3}(\tau_3)\rangle\sim e^{-iS}\,,
\end{equation}
where $S=\sum_{i=1,2,3}S_{\Delta_i}(\tau_i,u_I)$, being $S_{\Delta_i}(\tau_i,u_I)$ the contribution of an ingoing/returning geodesics from $\tau_i$ at the boundary to the intersection point $u_I$ --~to be fixed through extremization of $S$.

With no loss of generality, one can set $\tau_3=0$.
Then the relevant geodesic configuration is dynamically fixed and it depends on the values of $\tau_1, \, \tau_2$
and $\Delta_{1,2,3}$ (more specifically, on two ratios, {\it e.g.} $\Delta_1/\Delta_3$,  $\Delta_2/\Delta_3$). The key equation is

\begin{equation}
  \epsilon_1 \Delta_1 \sqrt{f(z_I)-\mu_1^2\,z_I^2}+
   \epsilon_2 \Delta_2 \sqrt{f(z_I)-\mu_2^2\,z_I^2}+ \epsilon_3 \Delta_3 \sqrt{f(z_I)-\mu_3^2\,z_I^2} =0\, ,
  \label{genat}
\end{equation}
with  $\Delta_1\,\mu_1+\Delta\,\mu_2+\Delta_3\,\mu_3=0$.
For given values of the parameters, this equation will have a unique real solution
$z_I\in (0,1)$, for a specific choice of signs $\epsilon_{1,2,3}$
(and other choices of sign will give no real solution $z_I\in (0,1)$).
This determines which geodesics are ingoing and which ones are returning.
For example, for $\tau_1$ sufficiently close to $\tau_2$,
the geodesics for operators 1 and 2 must be ingoing and the geodesic 3 must be returning. On the other hand, if we choose
$\tau_3=0$, $\tau_1$ and $\tau_2$ sufficiently separated and $\Delta_3<\Delta_1+\Delta_2$, geodesics 1 and 2 are returning and 3 is outgoing.
This will be illustrated by several examples in the following sections.

\medskip

\begin{figure}[h!]
\centering
\includegraphics[width=.6\textwidth]{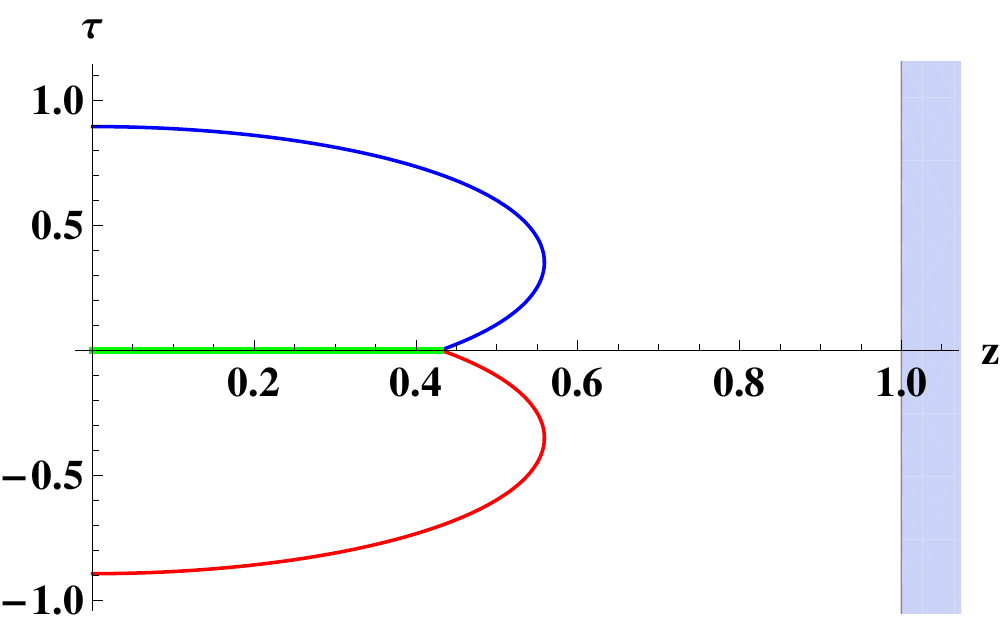}
\caption{A symmetric configuration for $\Delta_1=\Delta_2=30$ and $\Delta_3=40$, with the operator $O_3$ placed in the middle point between $O_1$ and $O_2$. Geodesics
1, 2 are returning and   geodesic 3 is ingoing. In this symmetric case, the latter is a  straight line between $z=0$ and $z_I$. [Red, blue and green colors represent the geodesics representing operators $O_1$, $O_2$ and $O_3$, respectively].
}
 \label{twosss}
\end{figure}

We begin with a case where calculations dramatically simplify. 
This is the case of a ``symmetric" correlator $\langle O_{\Delta}(\tau)\,O_{\Delta'}(0)\,O_{\Delta}(-\tau)\rangle$. In this case, the only consistent solution to the saddle-point equation implies that $O_{\Delta}(\tau)$ and $O_{\Delta}(-\tau)$  correspond to returning geodesics while $O_{\Delta'}(0)$  corresponds to an ingoing geodesic (see fig. \ref{twosss}). Then

\begin{eqnarray}
    &&\langle O_{\Delta}(\tau)\,O_{\Delta'}(0)\,O_{\Delta}(-\tau)\rangle \approx e^{-iS}\,,\nonumber\\ \nonumber\\
    &&S=S_{\Delta}^{\rm returning}(-\tau_1,u_I)+S_{\Delta}^{\rm returning}(\tau_1,u_I)+S_{\Delta'}^{\rm ingoing}(0,u_I)\,.
\end{eqnarray}
By symmetry, $\tau_I=0$, therefore the ingoing geodesic corresponding to $O_{\Delta'}(0)$ is actually a straight line  at constant $\tau=0$ that extends from the boundary up to $z_I$. This fixes the corresponding $\mu'=0$, and, through the second equation in \eqref{saddleequations}, sets $\mu_1=-\mu_2\equiv \mu$. Therefore, we are left with the first equation, which becomes

\begin{equation}
\label{saddlesymmetric}
    2\Delta\,\sqrt{f(z_I)-\mu^2\,z_I^2}-\Delta'\,\sqrt{f(z_I)}=0\,.
\end{equation}

\subsection{Symmetric three-point function in $d=2$}

As a warm-up, we start considering the symmetric thermal three-point function in  the simpler $d=2$ case. We consider the configuration of fig. \eqref{twosss}.
In two dimensions, the integral defining the geodesic $\tau(z,\mu)$ gives simple logarithmic functions, which permits to find an explicit closed formula for
$\mu $ in terms of the boundary coordinate $\tau$.
The saddle-point equation for $z$ is solved by
\begin{equation}
    z_I= z_{\rm max} \, \frac{\sqrt{4\Delta^2-{\Delta'}^2}}{\sqrt{4\Delta^2-z_{\rm max}^2{\Delta'}^2}}\ ,\qquad z_{\rm max} =\frac{z_0}{\sqrt{1+z_0^2\mu^2}}\ .
\end{equation}
Substituting into the equation for the geodesic, at the intersection point we have
$0 = \tau(z_I,\mu)$. This can be explicitly solved for $\mu$. We find ($z_0=1$)
\begin{equation}
    \mu = \cot \tau +\frac{\Delta'}{2\Delta \sin\tau} \ .
\nonumber
\end{equation}
Substituting into the action, we obtain

\begin{equation}
   \langle O_\Delta(\tau) O_{\Delta'}(0) O_{\Delta}(-\tau)\rangle= 
    \left(1-\frac{\Delta'}{2 \Delta }\right){}^{\Delta
   -\frac{\Delta'}{2}}
   \left(1+\frac{\Delta'}{2 \Delta
   }\right){}^{\Delta +\frac{\Delta'}{2}} \ 
\frac{   \cot ^{\Delta'}(\pi T\tau)}{ \sin ^{2 \Delta }(2\pi T \tau )}\ ,
\label{dddd}
\end{equation}
where the temperature has been restored, recalling that $z_0=1/(2\pi T)$.
This agrees with the result of \cite{Becker:2014jla}. Note that, in the ``extremal" limit $\Delta_0\to 2\Delta$, one obtains\footnote{By extremal correlators we mean those where the dimension of one operator equals the sum of the other two.}

\begin{equation}
    \langle O_\Delta(\tau) O_{2\Delta}(0) O_{\Delta}(-\tau)\rangle= 
   \frac{1}{(\sin (\pi T\tau ))^{4 \Delta }}\,,
\end{equation}
which is, up to a $\Delta$-dependent constant, the product of two 2-point functions, 

\begin{equation}
    \langle O_{\Delta}(\tau)\,O_{2\Delta}(0)\,O_{\Delta}(-\tau)\rangle=\mathcal{N}_{\Delta}\,\langle O_{\Delta}(\tau)\,O_{\Delta}(0)\rangle\,\langle O_{\Delta}(0)\,O_{\Delta}(-\tau)\rangle\,.
\end{equation}
We will return to this factorization property in sec. \ref{facsec}.

Beyond the extremal limit, when $\Delta' > 2 \Delta$, the saddle point no longer lies within the region $0<z<z_0$. The correlation function can be defined by analytic continuation and it is of the same form \eqref{dddd}
(for $2\Delta<  \Delta'< 2\Delta/z_{\rm max}$, the saddle point is imaginary; for $\Delta'> 2\Delta/z_{\rm max}$, it lies inside the horizon).

\subsection{Symmetric three-point function in $d=4$}

In $d=4$ the solution to \eqref{saddlesymmetric} is 
\begin{equation}
z_I=\sqrt{\frac{\sqrt{4(1-\delta^2)^2+\mu^2}-\mu^2}{2(1-\delta^2)}}\,,\qquad \delta=\frac{\Delta'}{2\Delta}\,.
\label{zai}
\end{equation}
Here we have chosen units where $1=z_0=\beta/\pi$. We assume $\delta<1$, that is, the `non-extremal case' $\Delta'<2\Delta$. As in the $d=2$ case,
for $\Delta'>2\Delta$ the correlator is defined by analytic continuation.

Substituting \eqref{zai}  into the geodesic equation $\tau=\tau(z,\mu)$, we find  the condition at the  intersection point, $0 = \tau(z_I,\mu)$.
This can be solved for $\mu$ in series of small $\tau_1$. We get

\begin{equation}
\mu=\frac{1+\frac{\Delta'}{2\Delta}}{\tau_1+\frac{32\Delta^2-18\Delta\Delta'+3\Delta'^2}{10\,(2\Delta+\Delta')^2}\,\tau_1^5+\cdots}\,.
\end{equation}
Substituting this into $S$ one finally finds 

\begin{eqnarray}
&&\langle O_{\Delta}(\tau_1)\,O_{\Delta'}(0)\,O_{\Delta}(-\tau_1)\rangle=\frac{e^{-iS_T}}{\tau_1^{2\Delta+\Delta'}}\,,\nonumber \\  \nonumber \\ &&  -i\,S_T=\frac{\left(32 \Delta ^2-18 \Delta \Delta' +3   \Delta'^2\right)}{40 (2 \Delta+\Delta')}\,\pi^4\,T^4\,\tau_1^4+\mathcal{O}(T^8\tau_1^8)\,.
\end{eqnarray}
where the temperature has been restored.
As a sanity check, in the $T=0$ limit we find $\langle O_{\Delta}(\tau_1)\,O_{\Delta'}(0)\,O_{\Delta}(-\tau_1)\rangle=\tau_1^{-(2\Delta+\Delta')}$, which is precisely the expected result (\textit{cf.} \eqref{3pointT=0}). 

\medskip

Let us now consider different limits. For the hierarchy $\Delta\gg \Delta'$, to leading order we have

\begin{equation}
\langle O_{\Delta}(-\tau_1)\,O_{\Delta'}(0)\,O_{\Delta}(-\tau_1)\rangle \approx \frac{e^{\frac{\pi^4}{40}\,\Delta\,T^4\,\tau^4}}{\tau^{2\Delta+\Delta'}}\,,
\end{equation}
where $\tau=2\tau_1$. This approaches the 2-point function $\langle O_{\Delta}(\tau_1)\,O_{\Delta}(-\tau_1)\rangle$. 

On the other hand, in the `extremal' limit $\Delta'=2\Delta$, we find

\begin{equation}
\langle O_{\Delta}(\tau_1)\,O_{2\Delta}(0)\,O_{\Delta}(-\tau_1)\rangle
\to \Big(\frac{e^{\frac{\pi^4\,\Delta\,T^4\,\tau_1^4}{40}}}{\tau_1^{2\Delta}}\Big)^2=\mathcal{N}_{\Delta}\,\langle O_{\Delta}(\tau_1)\,O_{\Delta}(0)\rangle \, \langle O_{\Delta}(0)\,O_{\Delta}(-\tau_1)\rangle\,,
\end{equation}
that is, the extremal 3-point function again factorizes as  the product of two 2-point functions.

\section{General three-point correlation function and factorization limit}\label{3-point}
\label{facsec}

\subsection{Factorization}

In the previous section, we considered the symmetric 3-point function $\langle O_{\Delta}(\tau_1)\,O_{\Delta'}(0)\,O_{\Delta}(-\tau_1)\rangle$ in $d=2,4$. The symmetric configuration makes it possible to find an analytic formula, which is exact in the $d=2$ case, and in the $d=4$ case it describes a $T|\tau|\ll 1$ regime. 
Both in $d=2,4$ the three-point function satisfies a  surprising factorization property in the ``extremal" limit
$\Delta'\to 2\Delta$.
%
In this subsection we will clarify the origin of this factorization and 
study it for general three-point functions.

At $T=0$, the factorization property is manifest  from the generic form of the 3-point function, which is entirely fixed by conformal symmetry,

\begin{equation}
\label{3pointT=0}
\langle O_{\Delta_1}(x_1)\,O_{\Delta_2}(x_2)\,O_{\Delta_3}(x_3)\rangle=\frac{f_{O_{\Delta_1}O_{\Delta_2}O_{\Delta_3}}}{|x_1-x_2|^{\Delta_1+\Delta_2-\Delta_3}\,|x_2-x_3|^{\Delta_2+\Delta_3-\Delta_1}\,|x_3-x_1|^{\Delta_3+\Delta_1-\Delta_2}}\,.
\end{equation}
Setting, for example, $\Delta_3=\Delta_1+\Delta_2$, one gets

\begin{eqnarray}
\langle O_{\Delta_1}(x_1)\,O_{\Delta_2}(x_2)\,O_{\Delta_1+\Delta_2}(x_3)\rangle &\sim&  \frac{1}{|x_1-x_3|^{2\Delta_1}\, |x_2-x_3|^{2\Delta_2}}\nonumber  \\ &\sim& \langle O_{\Delta_1}(x_1)\,O_{\Delta_1}(x_3)\rangle\,\langle O_{\Delta_2}(x_2)\,O_{\Delta_2}(x_3)\rangle \,. \nonumber
\end{eqnarray}
However, at $T\neq 0$ the form of the 3-point function is not fixed by symmetries and
it has an extremely complicated general structure.

To start with, 
let us first consider a generic three-point correlation function with only time-dependence. With no loss of generality, we can place one of the operators at $\tau=0$. 
Thus we consider correlators of the form $\langle O_{\Delta_1}(-\tau_1)\,O_{\Delta_3}(0 )\, O_{\Delta_2}(\tau_2)\rangle$ and, for concreteness, we will initially assume that with $\tau_1>0$ and $\tau_2>0$.
For large dimensions, $\Delta_i \gg 1$,  the correlation function is represented by three geodesic arcs meeting at some point $(\tau_I,z_I)$ in the bulk. An examination of the saddle-point equation shows that, for $\Delta_3< \Delta_1+\Delta_2$, and $\Delta_3$ sufficiently close to $ \Delta_1+\Delta_2$,  the operator $O_{\Delta_3}$ is described by an ingoing geodesic while $O_{\Delta_{1,2}}$ are described by  returning geodesics, generalizing the picture of fig. \ref{twosss}. 
The saddle point equations then read

\begin{eqnarray}
   && \Delta_1\, \sqrt{f(z_I)-\mu_1^2\,z_I^2}+\Delta_2\, \sqrt{f(z_I)-\mu_2^2\,z_I^2}-\Delta_3\, \sqrt{f(z_I)-\mu_3^2\,z_I^2}=0\, , \nonumber \\ \nonumber \\
  &&  \Delta_1\,\mu_1+\Delta_2\,\mu_2+\Delta_3\,\mu_3=0\,.
  \label{gentres}
\end{eqnarray}
Other choices of relative signs do not have real solution $0<z_I<1$ in the
region $\Delta_3= \Delta_1+\Delta_2-\epsilon$.
The saddle-point equation \eqref{gentres} can be solved explicitly in terms of one of the roots of the quartic polynomial for $X\equiv z^2$, obtained by squaring
the saddle-point equation.
For sufficiently small $\tau_1$ and $\tau_2$, as compared with the scale of the thermal circle $\beta$, there is a unique root that represents the dominant saddle point and describes the expected short-distance behavior of the correlation function.

In the region $\Delta_3 \sim \Delta_1 + \Delta_2$, it has the behavior
\begin{equation}
    z_I^2  \approx \frac{2}{\Delta_1 \mu_1^2+\Delta_2 \mu_2^2 -(\Delta_1+\Delta_2) \mu_3^2 }\ (\Delta_1 +\Delta_2 -\Delta_3) \approx 0
\end{equation}
Note that $z_I$ is real for $\Delta_3 \leq \Delta_1 + \Delta_2$, 
and imaginary for $\Delta_3 > \Delta_1 + \Delta_2$.

At the critical point  $\Delta_3 = \Delta_1 + \Delta_2$, we have $z_I= 0$, and the three point correlation function exhibits the remarkable factorization property:
\begin{eqnarray}
    \langle O_{\Delta_1}(-\tau_1)\,O_{\Delta_3}(0 )\, O_{\Delta_2}(\tau_2)\rangle &=& \langle O_{\Delta_1}(-\tau_1)\,O_{\Delta_1}(0 )\rangle \ \langle O_{\Delta_2}(0)\,O_{\Delta_2}(\tau_2 )\rangle 
        \nonumber\\
    \nonumber\\
    &\approx &  4^{-\Delta_1} \ \left(\mu_1^2+4\right)^\frac{\Delta_1}{2}  4^{-\Delta_2} \ \left(\mu_2^2+4\right)^\frac{\Delta_2}{2}\ ,
\end{eqnarray}
where $\mu_k(\tau_k)$, $k=1,2$, are implicitly defined by 
\begin{equation}
   \tau = -\frac12 \log \frac{\mu (\mu-2) +2}{\sqrt{\mu^2+4}}+ \frac{i}2 \log \frac{\sqrt{\mu^2+4}}{\mu (\mu+2i)-2}\ .
\end{equation}

\begin{figure}[h!]
\centering
\includegraphics[width=.45\textwidth]{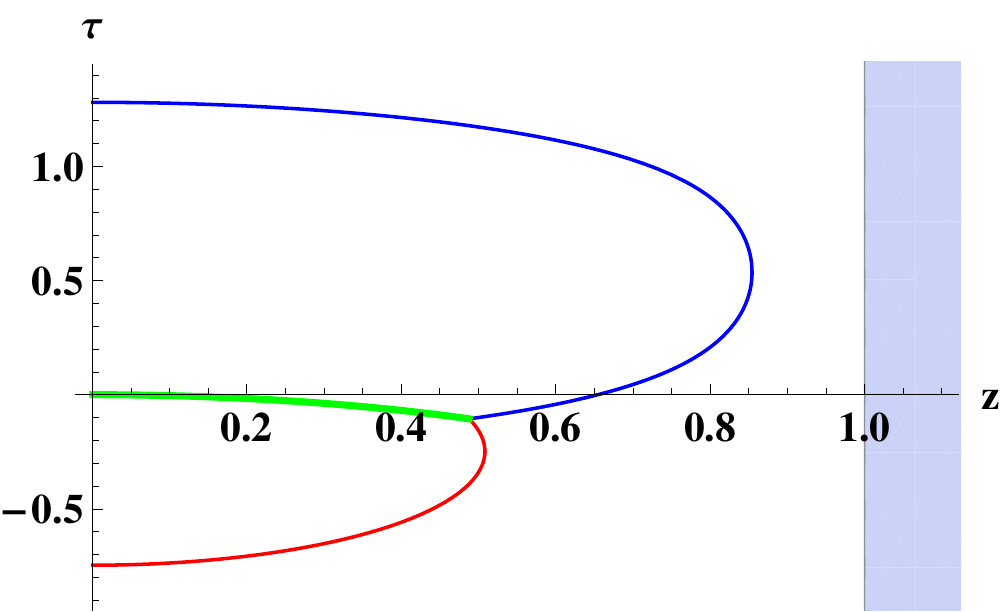}
\includegraphics[width=.45\textwidth]{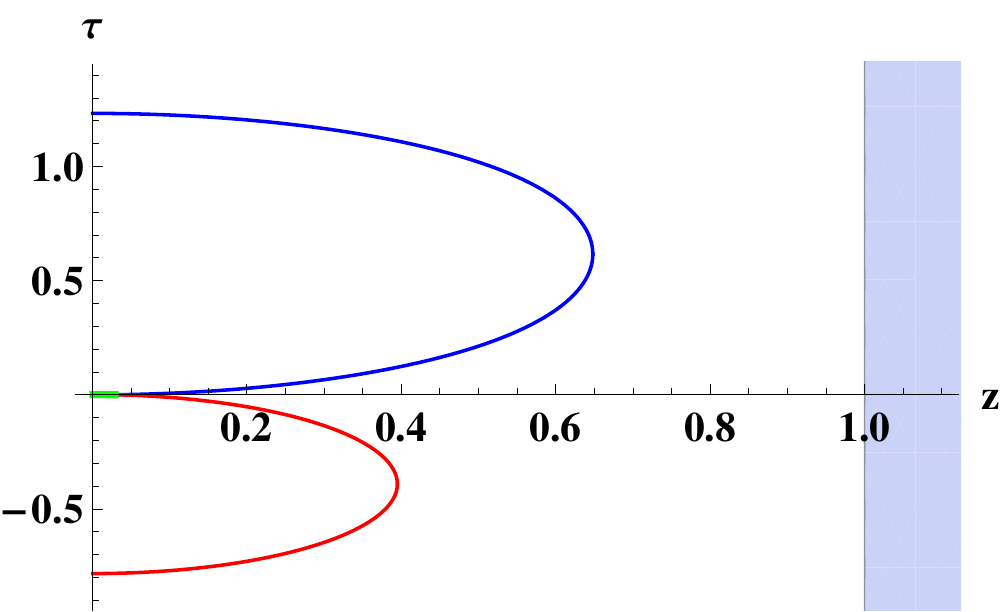}
\caption{a) Non-symmetric configuration for $\Delta_1=\Delta_2=30$ and $\Delta_3=40$, with the operator $O_{\Delta_3}$ placed between $O_{\Delta_1}$ and $O_{\Delta_2}$. b) As $\Delta_3\to \Delta_1+\Delta_2$, the configuration reaches the factorization limit. In the figure b, $\Delta_1=\Delta_2=30$ and $\Delta_3=59.9$. [Same color notation as in fig. \ref{twosss}]..
}
 \label{twosides}
\end{figure}

\begin{figure}[h!]
\centering
\includegraphics[width=.45\textwidth]{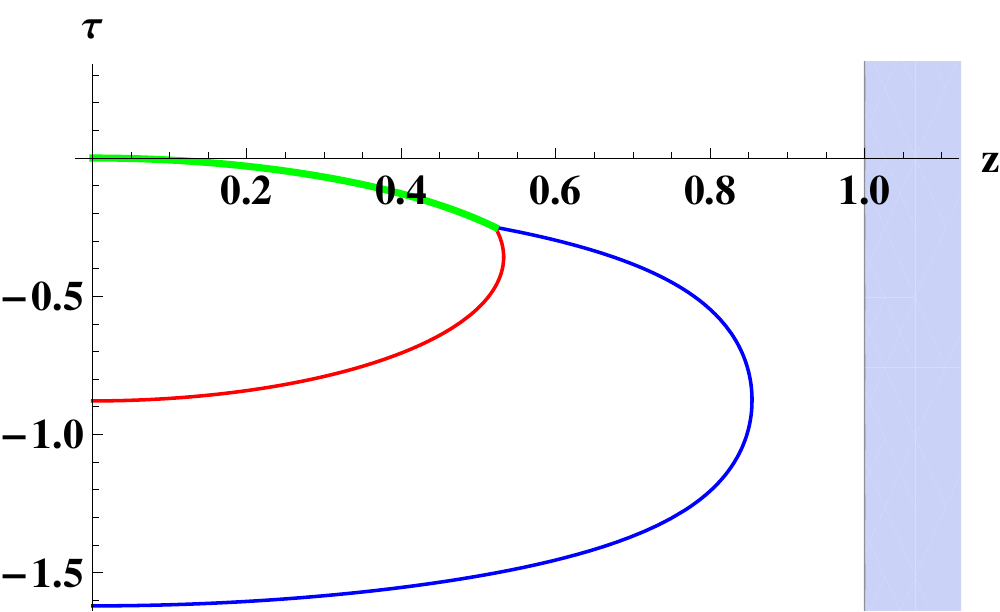}
\includegraphics[width=.45\textwidth]{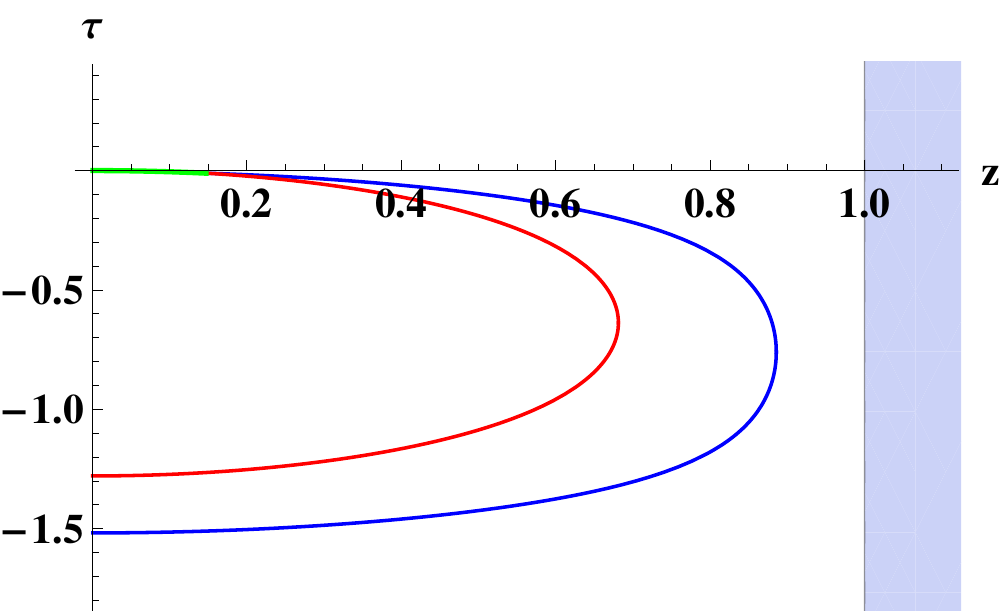}
\caption{a) Non-symmetric configuration for $\Delta_1=50$, $\Delta_2=50$ and $\Delta_3=90$, with the operator $O_{\Delta_1}$ placed between $O_{\Delta_2}$ and $O_{\Delta_3}$. b)  $\Delta_1=50$, $\Delta_2=50$ and $\Delta_3=99.9$. [Same color notation as in fig. \ref{twosss}].
}
 \label{onesidex}
\end{figure}

The factorization limit is illustrated in figs. \ref{twosides} and \ref{onesidex}.
One can see that the geodesic describing the operator $O_{\Delta_3}$ shrinks to zero
as $\Delta_3 \sim \Delta_1 + \Delta_2$. In this limit, the configuration reduces
to two individual geodesics ending at the same point $\tau_3=0$, which represent the 
two propagators $\langle O_{\Delta_1}(-\tau_1)\,O_{\Delta_1}(0 )\rangle $ and $  \langle O_{\Delta_2}(0)\,O_{\Delta_2}(\tau_2 )\rangle $.

\subsection{Adding $\vec x$ dependence}

Let us now incorporate the $\vec x$ dependence. The details are given in
appendix \ref{appendix:geodesics}.
In the general case, we  find the following saddle-point equations 

\begin{eqnarray}
   && \sum_{i=1}^3 \epsilon_i \Delta_i \sqrt{f(z_I)-\mu_i^2\,z_I^2+f(z_I)\nu_i^2 z_I^2 }=0\, , \nonumber \\ \nonumber \\
  &&  \Delta_1\,\mu_1+\Delta_2\,\mu_2+\Delta_3\,\mu_3=0\, ,
   \nonumber \\
  &&  \Delta_1\,\nu_1+\Delta_2\,\nu_2+\Delta_3\,\nu_3=0\,.
  \label{gentres}
\end{eqnarray}
where, again $\epsilon_i=\pm 1$ according to whether the corresponding geodesic
is ingoing or returning at the point $z_I$.

Let us assume that $O_1(\tau_1,\vec x_1)$ and $O_2(\tau_2,\vec x_2)$
are represented by the returning geodesics, and $O_3(\tau_3,\vec x_3)$
by an ingoing geodesic. 
The factorization limit then occurs when $z_I\to 0$. The saddle-point equation for $z_I$ then requires
\begin{equation}
    \sum_{i=1}^3 \epsilon_i \Delta_i = \Delta_3-\Delta_1-\Delta_2 \to 0\ .
\end{equation}
In such a case the ingoing geodesic shrinks to zero and one has 
\begin{equation}
\label{fact}
    \langle O_{\Delta_1}(x_1)\,O_{\Delta_2}(x_2)\,O_{\Delta_1+\Delta_2}(x_3)\rangle \to \mathcal{N}_{\Delta_1,\Delta_2}\,\langle O_{\Delta_1}(x_1)\,O_{\Delta_1}(x_3)\rangle\,\langle O_{\Delta_2}(x_2)\,O_{\Delta_2}(x_3)\rangle\,;
\end{equation}
Thus the factorization of the three-point function seems to be a general property,
which occurs in the limit   $\sum_{i=1}^3 \epsilon_i \Delta_i \to 0$.

\medskip

\begin{figure}[h!]
\centering
\includegraphics[width=.45\textwidth]{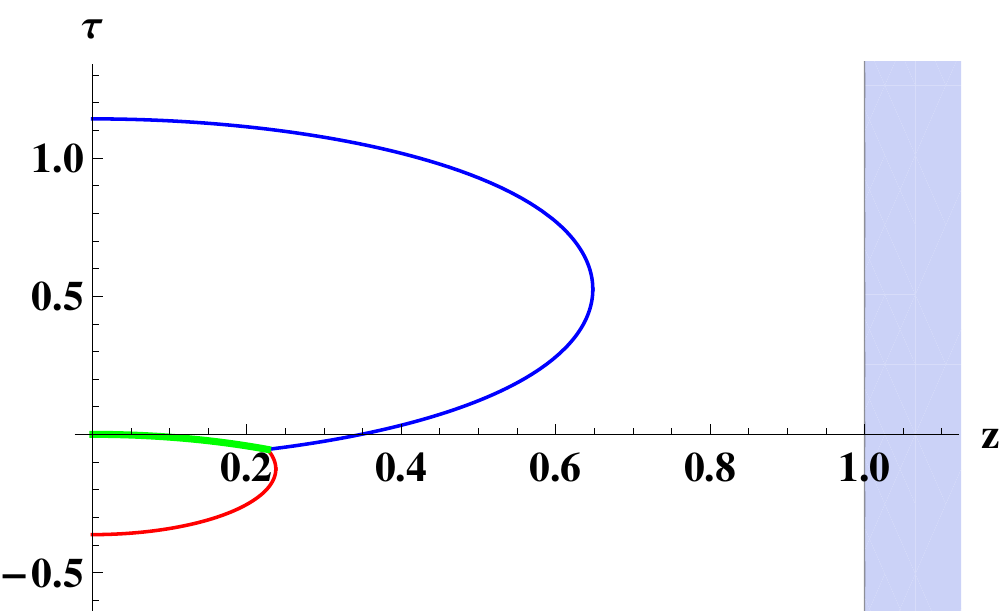}
\includegraphics[width=.45\textwidth]{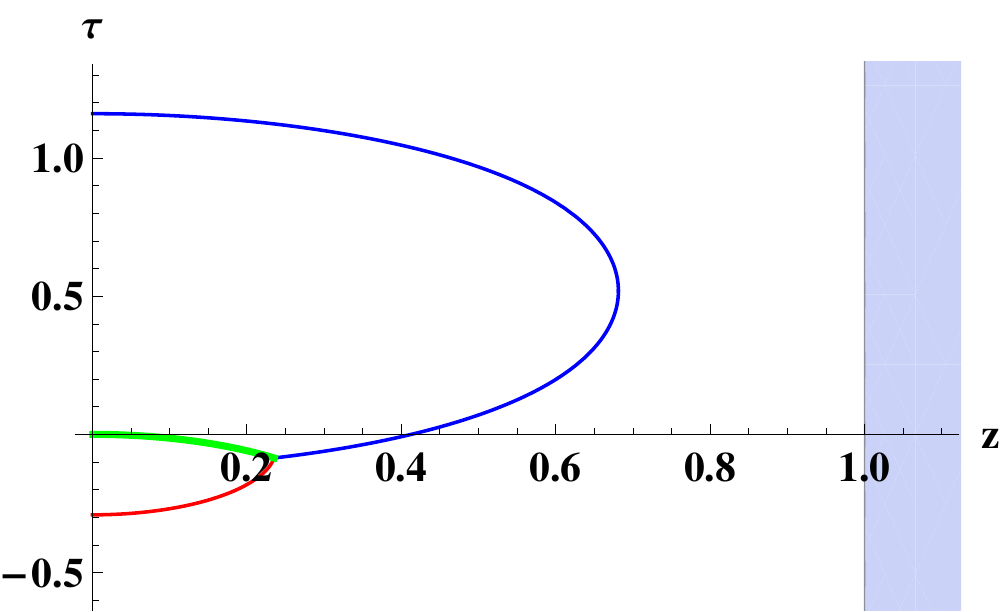}
\caption{a) Geodesic configuration for $\Delta_1=\Delta_2=30$, $\Delta_3=42$, $\tau_2=1.15$, $\tau_1=-0.37$.
b) Here $\Delta_1=\Delta_2=30$, $\Delta_3=32$, $\tau_2=1.15$, $\tau_1=-0.3$. The geodesic 1 (in red), which was  returning  in fig. a, is now ingoing. [Same color notation as in fig. \ref{twosss}]..
}
 \label{transiciones}
\end{figure}

As the location points of the operators move around in the boundary,
there can be transitions where a returning geodesic becomes ingoing, and vice-versa (see fig. \ref{transiciones}). A critical point occurs whenever the intersection point
coincides with one of the $z_{\rm max}$ of the three geodesics.
A simple examination shows that this never happens in the symmetric case
of section \ref{3-point-symmetric}. For example, the condition, $z_{\rm max}^{(1)}=z_I$ determines the critical value
$$
\mu_1^c=-\frac{2\Delta_1\Delta_2}{\Delta_1^2+\Delta_2^2-\Delta_3^2}\, \mu_2\ .
$$ 
Assuming $\mu_2<0$, the geodesic 1 is returning for $\mu_1<\mu_1^c$ and ingoing for $\mu_1>\mu_1^c$. 
For the symmetric case $\mu_1=-\mu_2$, $\Delta_1=\Delta_2$, and it is not possible to have a transition (geodesic 1 is always returning)

It would be very interesting to study this phenomenon in more detail,
in particular, the analytic properties of such transitions in the infinite $\Delta_i$ limit. 
On general grounds, one expects that
the three-point function should be analytic in $x_1$, $x_2$, $x_3$ away from the coinciding points. This is also supported by the fact that  ingoing-returning transitions also take place at $T=0$, with the same critical value $\mu_1^c$. In the $T=0$ case, the  well-known 3-point function has a simple form, as being completely fixed by conformal invariance. Clearly, it is an analytic function of $x_{1,2,3}$ away from the coinciding points.


In addition to ingoing-returning transitions, as the separation
of the operators along the thermal circle increases and the geodesics go further into the bulk, other saddle points have to be considered. This is analogous to what happens for 2-point functions in \cite{Fidkowski:2003nf,Rodriguez-Gomez:2021pfh}. 
Although in this paper we restricted the considerations to distances less than the critical distance corresponding to possible saddle-point jumps, it would clearly be very interesting to study these aspects in detail.

\subsection{Factorization property as a sign of a semiclassical limit}

At $T=0$, extremal correlators have been extensively discussed,  albeit mostly in connection with subtleties related to their prefactors: in top-down constructions from type IIB string theory \cite{Lee:1998bxa}, the overall coefficient typically vanishes while at the same time the coupling constant diverges in such a way that these effects cancel leaving a finite result \cite{Liu:1999kg,DHoker:1999jke}. As far as their spacetime dependence is concerned, because their form is completely fixed by conformal symmetry, factorization arises as a trivial consequence. It is striking that
the factorization property of extremal correlators  still seems to hold in the 
general $T\neq 0$ theory, where  three-point functions, as well as  two-point functions, have a much more complicated structure. 

To investigate the implications of this property, let us suppose that we
start with a general CFT and take, as a fact,  that extremal three-point (connected) correlation functions 
factorize as

\begin{equation}
\label{fact}
    \langle O_{\Delta_1}(x_1)\,O_{\Delta_2}(x_2)\,O_{\Delta_1+\Delta_2}(x_3)\rangle=\mathcal{N}_{\Delta_1,\Delta_2}\,\langle O_{\Delta_1}(x_1)\,O_{\Delta_1}(x_3)\rangle\,\langle O_{\Delta_2}(x_2)\,O_{\Delta_2}(x_3)\rangle\,;
\end{equation}
where all operators are assumed to be primaries. 
Consider the $x_2\rightarrow x_1$ limit. 
In this limit we can replace $O_{\Delta_1}(x_1)\,O_{\Delta_2}(x_2)$ by the OPE,
which would express the RHS of \eqref{fact} as a sum over two-point functions.

Let us make the extra assumption that two-point functions of non-identical operators vanish.\footnote{At $T\neq 0$ two-point functions of different operators may not vanish. Here we 
assume that  a diagonal basis exists.
This seems to be suggested by holography and by the infinite $\beta$  limit.
}
The relevant term of the OPE is
$$
O_{\Delta_1}(x_1)\,O_{\Delta_2}(x_2)= C_{\Delta_1\Delta_2} O_{\Delta_1+\Delta_2}(x_1)+\cdots 
$$
Then, substituting into \eqref{fact}, one obtains

\begin{equation}
\label{bootstrapeq}
 C_{\Delta_1\Delta_2} \langle O_{\Delta_1+\Delta_2}(x_1)\,O_{\Delta_1+\Delta_2}(x_3)\rangle=\mathcal{N}_{\Delta_1,\Delta_2}\,\langle O_{\Delta_1}(x_1)\,O_{\Delta_1}(x_3)\rangle\,\langle O_{\Delta_2}(x_1)\,O_{\Delta_2}(x_3)\rangle\,.
\end{equation}
Equation \eqref{bootstrapeq} is of the form $F(\Delta_1+\Delta_2)= F(\Delta_1)F(\Delta_2)$. This relation ``bootstraps" the 2-point function of primary operators of dimension $\Delta$ to be of the form

\begin{equation}
\label{generalform}
    \langle O_{\Delta}(x) O_{\Delta}(0)\rangle=\mathcal{C}_{\Delta}\,e^{-\Delta\,\ell(x)}\,.
\end{equation}
This is precisely the form of the 2-point function that arises from the geodesic approximation. Thus we see that the factorization property implicitly requires the semiclassical approximation of large $\Delta$. 
At finite $\Delta$, correlation functions do not have, in general, the form
\eqref{generalform}.
An exception is $d=2$, where the exact thermal two-point function is of the form \eqref{generalform}, in which case the factorization  \eqref{fact} also holds for operators of small dimension.
For $d>2$, the factorization property for extremal correlators does not generally hold beyond the large $\Delta$ approximation.

On general grounds, $\ell(x)$ must be of the form

\begin{equation}
\label{formael}
\ell(x)=\log|x|^2+\sum_{\delta=\delta_0} c_{\delta}\,(T|x|)^{\delta}\,f_{\delta}(x)\,,
\end{equation}
where $f_{\delta}(x)$ are dimensionless $SO(d-1)$ invariant functions of $\eta=\frac{\tau}{\sqrt{\tau^2+\vec{x}^2}}$.  Since $\eta\in [-1,1]$, the $f_{\delta}$ can naturally be expressed in terms of  Gegenbauer polynomials, which are orthogonal functions in $[-1,1]$. In this way one recovers the general form discussed in \cite{Iliesiu:2018fao}. A consequence of the general form \eqref{generalform} is that the double scaling limit of \cite{Rodriguez-Gomez:2021pfh}, $T|x|\rightarrow 0$, $\Delta\rightarrow \infty$, with fixed $\Delta (T|x|)^{\delta_0}$,   exists and is well defined
(in the strong coupling expansion, one has $\delta_0=d$
as in \eqref{thermallimit}).

\section*{Acknowledgements}

We thank A. Tseytlin for  useful remarks.
D.R-G is partially supported by the Spanish government grant MINECO-16-FPA2015-63667-P. He also acknowledges support from the Principado de Asturias through the grant FC-GRUPIN-IDI/2018/000174 J.G.R. acknowledges financial support from projects 2017-SGR-929, MINECO
grant PID2019-105614GB-C21, and  from the State Agency for Research of the Spanish Ministry of Science and Innovation through the “Unit of Excellence María de Maeztu 2020-2023” (CEX2019-000918-M).

\begin{appendix}

\section{Details on the geodesic arcs}\label{appendix:geodesics}

In this appendix we study each type of geodesic, with particular care to their contribution to the saddle point equations. In this appendix we consider general geodesics with both $\vec{x}$ and $t$ dependence.

Let us consider a geodesic departing from the boundary of the form $t=t(z)$, $\vec x=\vec x(z)$.
 The action is 

\begin{equation}
S=-m\,\int dz \frac{1}{z}\,\sqrt{f\dot{t}^2-\frac{1}{f}-\dot{\vec{x}}^2}\,.
\end{equation}
The conjugate momenta to $t$ and $\vec  x$
(denoted, respectively, by $P_t\equiv m\mu$ and $\vec P\equiv m \vec \nu$)
are conserved. We choose a  coordinate system where
$\vec P = (P_x,0,0)$, so the motion is in the subspace $(z,t,x,0,0)$.
It is convenient to introduce the euclidean time $\tau = i t$.
Using momentum conservation, we can formally integrate the equations of motion, 

\begin{equation}
\label{eom}
\tau=\tau_1+I_{\tau}(z)\,,\qquad x=x_1+I_x(z)\, ;
\end{equation}
where $\tau_1$ and $x_1$ represent the boundary values and

\begin{equation}
I_{\tau}(z)=\int_0^z dz\,\frac{z\,\mu}{f\,\sqrt{f-z^2\mu^2+fz^2\nu^2}}\,,\qquad I_x(z)=  i \int_0^z dz\,\frac{z\,\nu}{\sqrt{f-z^2\mu^2+fz^2\nu^2}}\,.
\end{equation}
Note that trajectories that depart from the boundary have imaginary time $t$, therefore
real euclidean time $\tau=i t$. As well known, geodesics of physical particles moving in AdS   space with real time $t$ and real $\vec x$ do not get to the boundary. However, here we are interested in the above ``unphysical" geodesics departing from the boundary, which are those that describe the Green's function, as can be shown using the WKB approximation to the Green's equation.

Substituting the solution into the action and using that $m\approx \Delta$, we get

\begin{equation}
S=-i \Delta\,\int dz\,\frac{1}{z\sqrt{f-z^2\mu^2+fz^2\nu^2}}\,.
\end{equation}

Geodesics departing from the boundary reach a maximum value $z_{\rm max}$ 
and then they return to the boundary, following  a ``U" shape trajectory.
$z_{\rm max}$ is  a solution of 
$$
f(z_{\rm max})-z_{\rm max}^2\mu^2+f(z_{\rm max})z_{\rm max}^2\nu^2=0
$$
and it is
the point where $\dot \tau $ and $\dot x $ are infinity.
In computing the three-point function, we shall consider three geodesic arcs
meeting at some intersection point $(z_I,\tau_I,x_I)$.
Each arc may meet the intersection point before reaching $z_{\rm max}$, or
in their return way. We will refer to the first type of geodesic as {\it ingoing} geodesics, whereas the second type will be called {\it returning} geodesics.
This distinction is important, because a solution of the saddle-point equations
with $0<z_I<1$ requires that there must be at least one arc of each type.

For  ingoing geodesics, the corresponding action is obtained by integrating from the boundary to the intersecting point

\begin{equation}
\label{Singoing}
S^{\rm ingoing}= -i\Delta\,\int_\epsilon ^{z_I} dz\,\frac{1}{z\sqrt{f-z^2\mu^2+fz^2\nu^2}} \,   .
\end{equation}
From the equations for the trajectory, one gets a relation between $z_I$ and $\tau_I, x_I$:

\begin{equation}
\label{eomingoing}
\tau^{\rm ingoing}_I=\tau_1+I^{\rm ingoing}_{\tau}\,,\quad x_I^{\rm ingoing}=x_1+I^{\rm ingoing}_x\,;\quad I^{\rm ingoing}_{\tau}=I_{\tau}(z_I)\,,\ \  I^{\rm ingoing}_x=I_x(z_I)\,.
\end{equation}

The action for returning geodesics can be computed by considering the action of  a  complete geodesic departing the boundary, entering the bulk and coming back to the boundary, subtracting   the last piece from $z_I$ to the boundary. Thus, we can write

\begin{equation}
S^{\rm returning}=2S^{\rm max}-S^{\rm ingoing}\,,
\end{equation}
and likewise

\begin{equation}
\tau=2\tau^{\rm max}-\tau^{\rm ingoing}\,,\qquad x=2x^{\rm max}-x^{\rm ingoing}\, ,
\end{equation}
where 
\begin{equation}
\label{max}
S^{\rm max}=-i \Delta\,\int_\epsilon^{z_{\rm max}}dz\,\frac{1}{z\sqrt{f-z^2\mu^2+fz^2\nu^2}} \,,
\end{equation}
\begin{equation}
\qquad \tau^{\rm max}=\tau_1+I_{\tau}(z_{\rm max})\,,\qquad  x_I^{\rm max}=x_1+I_x(z_{\rm max})\,.
\end{equation}

The intersection point is obtained by solving the saddle-point equations
\begin{equation}
    \frac{d S}{d z_I}=0\ ,\qquad  \frac{d S}{d \tau_I}=0\ ,\qquad \frac{d S}{d x_I}=0\ .
\end{equation}
In computing these derivatives, we must take into account the fact that
$\mu $ and $\nu$ are functions of $z_I, \tau_I, x_I$ (and of $\tau_1$, $x_1$)
defined by the conditions \eqref{eomingoing}.

We will need to use the following identities:
\begin{eqnarray}
\label{identitiesS1}
&& \frac{\partial S^{\rm ingoing/returning}}{\partial \mu}=-i\Delta\mu\frac{\partial I^{\rm ingoing/returning}_{\tau}}{\partial \mu}+\Delta\nu\frac{\partial I^{\rm ingoing/returning}_x}{\partial\mu}\,,\\ &&
\label{identitiesS2} \frac{\partial S^{\rm ingoing/returning}}{\partial \nu}=-i\Delta\mu\frac{\partial I^{\rm ingoing/returning}_{\tau}}{\partial \nu}+\Delta\nu\frac{\partial I^{\rm ingoing/returning}_x}{\partial\nu}\,.
\end{eqnarray}
In addition, by taking the differential of \eqref{eomingoing} (or of the analog equations  for the returning case) one obtains the following ``constitutive" relations,

\begin{align}
&  \frac{\partial I_{\tau}}{\partial\mu}\frac{\partial\mu}{\partial \tau_I}+\frac{\partial I_{\tau}}{\partial\nu}\frac{\partial\nu}{\partial \tau_I}=1\,, &  \frac{\partial I_{x}}{\partial\mu}\frac{\partial\mu}{\partial \tau_I}+\frac{\partial I_{x}}{\partial\nu}\frac{\partial\nu}{\partial \tau_I}=0\,,\label{line1} \\
& \frac{\partial I_{\tau}}{\partial\mu}\frac{\partial\mu}{\partial x_I}+\frac{\partial I_{\tau}}{\partial\nu}\frac{\partial\nu}{\partial x_I}=0\,,& \frac{\partial I_{x}}{\partial\mu}\frac{\partial\mu}{\partial x_I}+\frac{\partial I_{x}}{\partial\nu}\frac{\partial\nu}{\partial x_I}=1\,, \label{line2}\\ 
&  \frac{\partial I_{\tau}}{\partial z_I}+\frac{\partial I_{\tau}}{\partial\mu}\frac{\partial\mu}{\partial z_I}+\frac{\partial I_{\tau}}{\partial\nu}\frac{\partial\nu}{\partial z_I}=0\,, & \frac{\partial I_{x}}{\partial z_I}+\frac{\partial I_{x}}{\partial\mu}\frac{\partial\mu}{\partial z_I}+\frac{\partial I_{x}}{\partial\nu}\frac{\partial\nu}{\partial z_I}=0\,. \label{line3}
\end{align}

\subsection{The $z_I$ saddle-point equation}

The contribution to the saddle-point equation in $z_I$ is

\begin{equation}
\label{zsaddle}
\frac{dS}{dz_I}=\frac{\partial S}{\partial z_I}+\frac{\partial S}{\partial\mu}\frac{\partial \mu}{\partial z_I}+\frac{\partial S}{\partial\nu}\frac{\partial \nu}{\partial z_I}\,.
\end{equation}
Using  \eqref{line3}, one gets

\begin{equation}
\label{zsaddle2}
\frac{dS}{dz_I}=\frac{\partial S}{\partial z_I}+ \frac{ \frac{\partial S}{\partial \mu} \,\frac{\partial I_{x}}{\partial \nu} - \frac{\partial S}{\partial \nu}\,\frac{\partial I_{x}}{\partial \mu}}{\frac{\partial I_{x}}{\partial \mu}\, \frac{\partial I_{\tau}}{\partial \nu}-\frac{\partial I_{\tau}}{\partial \mu}\, \frac{\partial I_{x}}{\partial \nu}}\,\frac{\partial I_{\tau}}{\partial z_I}+ \frac{ \frac{\partial S}{\partial \nu} \,\frac{\partial I_{\tau}}{\partial \mu} - \frac{\partial S}{\partial \mu}\,\frac{\partial I_{\tau}}{\partial \nu}}{\frac{\partial I_{x}}{\partial \mu}\, \frac{\partial I_{\tau}}{\partial \nu}-\frac{\partial I_{\tau}}{\partial \mu}\, \frac{\partial I_{x}}{\partial \nu}}\,\frac{\partial I_{x}}{\partial z_I}\,.
\end{equation} 
Using now \eqref{identitiesS1}, \eqref{identitiesS2}, this becomes

\begin{equation}
\label{zsaddle3}
\frac{dS}{dz_I}=\frac{\partial S}{\partial z_I}+i\Delta\mu\,\frac{\partial I_{\tau}}{\partial z_I}-\Delta\nu \,\frac{\partial I_{x}}{\partial z_I}\,.
\end{equation} 
Using now the explicit expressions of $S$, $I_\tau$ and $I_x$, we finally get

\begin{equation}
\label{zsaddle4}
\frac{dS}{dz_I}=-\frac{ i \epsilon\Delta }{z_I\,f(z_I)} \ \sqrt{f(z_I)-z_I^2\mu^2+f(z_I)z_I^2\nu^2}\,,
\end{equation} 
where $\epsilon=1$ for ingoing geodesics and  $\epsilon=-1$  for returning
geodesics.
The complete saddle-point equation for $z_I$ is   obtained by adding the contributions 
 \eqref{zsaddle4} from the different geodesic arcs.

\subsection{The saddle-point equations for $x_I$ and $\tau_I$}

Since $S$ has no explicit $x_I$, $\tau_I$ dependence, we have

\begin{equation}
\frac{dS}{dX}=\frac{\partial S}{\partial \mu}\frac{\partial \mu}{\partial X}+\frac{\partial S}{\partial \nu}\frac{\partial \nu}{\partial X}\,,
\end{equation}
where $X$ is either $x_I$ or $\tau_I$. Using   \eqref{line1} and \eqref{line2}, one finds

\begin{equation}
\frac{dS}{d\tau_I}=\frac{\frac{\partial I_{x}}{\partial \mu} \, \frac{\partial S}{\partial \nu} - \frac{\partial I_{x}}{\partial \nu}\,\frac{\partial S}{\partial\mu}}{\frac{\partial I_{x}}{\partial \mu}\,\frac{\partial I_{\tau}}{\partial \nu}-\frac{\partial I_{\tau}}{\partial \mu}\,\frac{\partial I_{x}}{\partial \nu}} \,,\qquad \frac{dS}{dx_I}=\frac{\frac{\partial I_{\tau}}{\partial \nu} \, \frac{\partial S}{\partial \mu} - \frac{\partial I_{\tau}}{\partial \mu}\,\frac{\partial S}{\partial\nu}}{\frac{\partial I_{x}}{\partial \mu}\,\frac{\partial I_{\tau}}{\partial \nu}-\frac{\partial I_{\tau}}{\partial \mu}\,\frac{\partial I_{x}}{\partial \nu}} \,.
\end{equation}
Finally, using the identities in \eqref{identitiesS1}, \eqref{identitiesS2}, we obtain

\begin{equation}
\label{xxxtt}
\frac{dS}{d\tau_I}=-i\Delta\,\mu\,,\qquad \frac{dS}{dx_I}=\Delta\,\nu\,.
\end{equation}
By adding the contributions \eqref{xxxtt} for the different geodesic arcs,
we obtain the general saddle-point equations \eqref{gentres} anticipated in section 5.2.

\end{appendix}

\end{document}